\documentclass[letterpaper]{jpconf}
\usepackage{graphicx}
\newcommand{\ba}{\begin{eqnarray}}
\newcommand{\ea}{\end{eqnarray}}

\begin{document}

\title{Valence and sea quarks in the nucleon}

\author{Roelof Bijker}

\address{Instituto de Ciencias Nucleares, 
Universidad Nacional Aut\'onoma de M\'exico, 
A.P. 70-543, 04510 M\'exico, D.F., M\'exico}

\ead{bijker@nucleares.unam.mx}

\author{Elena Santopinto}

\address{I.N.F.N., Sezione di Genova, via Dodecaneso 33, Genova, I-16146 Italy}

\ead{elena.santopinto@ge.infn.it}

\begin{abstract}
In this contribution, we discuss the spin and flavor content of the proton in the framework 
of the unquenched quark model, and address the role of valence and sea quarks in the nucleon. 
\end{abstract}

\section{Introduction}

The role of valence and sea quarks in the nucleon is addressed in the framework of the 
unquenched quark model. The constituent quark model (CQM) describes the nucleon as a system 
of three constituent, or valence, quarks. Despite the successes of the CQM ({\it e.g.} masses, 
electromagnetic couplings, magnetic moments), there is compelling evidence for the presence of 
sea quarks from the measurement of the flavor asymmetry of the proton and the so-called proton 
spin crisis. The role of the pion cloud in the nucleon has been the subject of many studies 
\cite{Kumano,Speth,GarveyPeng}, and was shown to hold the key to understand the flavor 
asymmetry and the spin-crisis of the proton. Recently, it was pointed out these two 
properties are closely related: angular momentum conservation of the pionic fluctuations 
of the nucleon leads to a relation between the flavor asymmetry and the contribution of 
orbital angular momentum to the spin of the proton ${\cal A}(p) = \Delta L$ \cite{Garvey}. 
This identity can be understood from the fact that the flavor asymmetry is a matrix element 
in isospin space, and the orbital angular momentum in spin space with the same values of the 
quantum numbers. 

The aim of this contribution is to study the properties of the nucleon in the unquenched quark 
model (UQM) at the level of a toy model in which only the effects of the pion cloud is taken into 
account. It is shown that the pion cloud offers a qualitative understanding of the results 
obtained in previous numerical studies \cite{uqm}, and thus provides important insights into the 
properties of the nucleon. 

\section{Flavor and spin content}

In the unquenched quark model the effect of the quark-antiquark pairs is taken into account via a 
$^{3}P_{0}$ creation mechanism. The resulting baryon wave function is given by \cite{uqm} 
\ba 
\left| \psi_A \right> = {\cal N} \left[ \left| A \right>  
+ \sum_{BC l J} \int d \vec{K} k^2 dk \, \left| BC,l,J; \vec{K},k \right> \, 
\frac{ \left< BC,l,J; \vec{K},k \left| T^{\dagger} \right| A \right> } 
{\Delta E_{BC}(k)} \right] ~, 
\label{wf1}
\ea
where $\Delta E_{BC}(k) = M_A - E_B(k) - E_C(k)$ is the energy difference calculated in the rest 
frame of the initial baryon $A$ with $E_B(k)=\sqrt{M_B^2+k^2}$ and $E_C(k)=\sqrt{M_C^2+k^2}$. 
The operator $T^{\dagger}$ is the $^{3}P_0$ quark-antiquark pair creation operator 
\cite{uqm,Roberts}; $\vec{k}$ and $l$ denote the relative radial momentum and orbital 
angular momentum of $B$ and $C$, and $J$ is the total angular momentum 
$\vec{J} = \vec{J}_B + \vec{J}_C + \vec{l}$. The strength of the $^{3}P_{0}$ coupling 
is determined from the flavor asymmetry of the proton. 

In this contribution, we employ a simplified version of the UQM in which only the contribution  
of the pion cloud is taken into account. Table~\ref{quarkmodels} shows the results for the flavor 
and spin content of the proton. In the UQM, the three coefficients $a^2$, $b^2$ and $ab$ are expressed 
in terms of an integral over the relative momentum $k$ which depends on the $^{3}P_{0}$ coupling strength.
We note, that the results for the UQM in Table~\ref{quarkmodels} also hold for the meson-cloud model in 
which the coefficients $a$ and $b$ multiply the $N \pi$ and $\Delta \pi$ components of the nucleon 
wave function. The $ab$ term denotes the contribution from the cross terms between the $N \pi$ and 
$\Delta \pi$ components. In the UQM the value of the cross term $ab$ is not equal to the product of $a$ 
and $b$, although it turns out that the numerical values are close. 

\begin{table}
\centering
\caption{\small Spin and flavor content of the proton in he constituyent quark model (CQM) 
and the unquenched quark model (UQM).}
\label{quarkmodels}
\vspace{15pt}
\begin{tabular}{ccc}
\noalign{\smallskip}
\hline
\noalign{\smallskip}
\hline
\noalign{\smallskip}
& CQM & UQM \\
\noalign{\smallskip}
\hline
\noalign{\smallskip}
${\cal A}(p)=\Delta L$ & $0$ & $\frac{2a^2-b^2}{3(1+a^2+b^2)}$ \\
\noalign{\smallskip}
$\Delta u$ & $ \frac{4}{3}$ & $ \frac{4}{3}-\frac{38a^2+b^2-16ab\sqrt{2}}{27(1+a^2+b^2)}$ \\ 
\noalign{\smallskip}
$\Delta d$ & $-\frac{1}{3}$ & $-\frac{1}{3}+\frac{2a^2+19b^2-16ab\sqrt{2}}{27(1+a^2+b^2)}$ \\
\noalign{\smallskip}
$\Delta s$ & $0$ & $ 0$ \\
\noalign{\smallskip}
$\Delta \Sigma = \Delta u + \Delta d + \Delta s$ & $1$ & $1-\frac{4a^2-2b^2}{3(1+a^2+b^2)}$ \\
\noalign{\smallskip}
$g_A = \Delta u - \Delta d$ & $\frac{5}{3}$ & $\frac{5}{3}-\frac{40a^2+20b^2-32ab\sqrt{2}}{27(1+a^2+b^2)}$ \\
\noalign{\smallskip}
\hline
\noalign{\smallskip}
\hline
\end{tabular}
\end{table}

Since the UQM contains the full spin and isospin structure, it satisfies the 
relation between the flavor asymmetry and the contribution of the orbital angular momentum to the 
spin of the proton ${\cal A}(p)=\Delta L$ \cite{Garvey}, and therefore $\Delta \Sigma = 1-2\Delta L$. 
This relation does not hold for the chiral quark model of \cite{Eichten,ChengLi} in which the orbital 
angular momentum is enhanced with respect to the flavor asymmetry $\Delta L = 3{\cal A}(p)/2$ as a 
consequence of the requirement of a helicity flip of the quark. 

\begin{table}[ht]
\centering
\caption{\small Spin and flavor content of the proton normalized to the flavor asymmetry, 
UQM1 using the E866/NuSea value \cite{Towell} and UQM2 using the NMC value \cite{NMC}.}
\label{QM}
\vspace{15pt}
\begin{tabular}{ccrrcc}
\noalign{\smallskip}
\hline
\noalign{\smallskip}
\hline
\noalign{\smallskip}
& CQM & UQM1 & UQM2 & Exp & Ref \\
\noalign{\smallskip}
\hline
\noalign{\smallskip}
${\cal A}(p)$   & $0$    & $*0.118$ & $*0.158$ & $ 0.118 \pm 0.012$ & \cite{Towell} \\
&&&& $0.158 \pm 0.010$ & \cite{NMC} \\
\noalign{\smallskip}
$\Delta u$      & $ 4/3$ & $ 1.132$ & $ 1.064$ & $ 0.842 \pm 0.013$ & \cite{Hermes} \\ 
\noalign{\smallskip}
$\Delta d$      & $-1/3$ & $-0.368$ & $-0.380$ & $-0.427 \pm 0.013$ & \cite{Hermes} \\
\noalign{\smallskip}
$\Delta s$      & $0$ & $0$ & $0$ & $-0.085 \pm 0.018$ & \cite{Hermes} \\
\noalign{\smallskip}
$\Delta \Sigma$ & $1$    & $ 0.764$ & $ 0.684$ & $0.330 \pm 0.039$ & \cite{Hermes} \\
\noalign{\smallskip}
$g_A$           & $ 5/3$ & $ 1.500$ & $ 1.444$ & $1.2701 \pm 0.0025$ & \cite{PDG} \\
\noalign{\smallskip}
\hline
\noalign{\smallskip}
\hline
\end{tabular}
\end{table}

Table~\ref{QM} shows the results for the spin and flavor content of the proton normalized to 
the proton flavor asymmetry. The third column is normalized to the E866/NuSea value 
\cite{Towell}, and the fourth column to the somewhat higher NMC value \cite{NMC}. 
The experimental values of the spin content were obtained by the HERMES \cite{Hermes} and the 
COMPASS \cite{Compass} Collaborations. In Table~\ref{QM}, we show the HERMES results. 

The probability that a proton fluctuates in $n \pi^+$ 
\ba
\left| \left< n \pi^+ | p \right> \right|^2 \;=\; \frac{2a^2}{3(1+a^2+b^2)} \;=\; 0.180 ~,
\ea
(UQM1 value) is in close agreement with the experimental value $0.17 \pm 0.01$ determined in an 
analysis of forward neutron production in electron-proton collisions at 300 GeV by the H1 and 
ZEUS Collaborations at DESY \cite{Povh,Rosina1}. The UQM2 value is somewhat higher 0.241.
The total probability for a pion fluctuation of the proton is given by   
\ba
\left| \left< N \pi | p \right> \right|^2 + \left| \left< \Delta \pi | p \right> \right|^2 
\;=\; \frac{a^2+b^2}{1+a^2+b^2} \;=\; 0.455 ~,
\ea
(UQM1 value), in good agreement with the value of $0.470$ as determined in an analysis of the 
quark distribution functions measured in Drell-Yan experiments and semi-inclusive DIS experiments 
\cite{Chang}. Also in this case, the UQM2 value, $0.609$, is about 30 \% higher than the UQM1 value.  

\section{Summary and conclusions}

In this contribution, we studied the properties of the proton in the framework of the unquenched quark 
model in which the $^{3}P_0$ coupling strength was normalized to the observed value of the proton 
flavor asymmetry. It was shown that the pion fluctuations help to understand the discrepancies between 
the constituent quark model and the experimental data. Their inclusion leads to a 
reduction of quark model value of $\Delta u$ and $g_A$, and give rise to a sizeable 
contribution (25 - 30 \%) of orbital angular momentum to the spin of the proton. In addition, 
it was found that the probabilities for pion fluctuations in the UQM are in good agreement 
with the values determined in analyses of the available experimental data. 

\ack
This work was supported in part by research grants from CONACyT and PAPIIT-UNAM.

\end{document}